\UseRawInputEncoding
\documentclass[aps,prc,twocolumn,showkeys,showpacs]{revtex4}
\usepackage{graphicx}
\usepackage{dcolumn}
\usepackage{color}
\usepackage{bm}

\begin{document}

\newcommand{\req}[1]{(\ref{#1})}
\newcommand{\bel}[1]{\begin{equation}\label{#1}}
\newcommand{\belar}[1]{\begin{eqnarray}\label{#1}}
\def\bra{\langle}
\def\ket{\rangle}
\def\eps{\epsilon}
\def\epsk{\epsilon_k}
\def\mev{\;{\rm MeV}}
\def\tc{\textcolor{red}}
\def\tb{\textcolor{}}
\def\pr{\prime}
\def\dpr{\prime\prime}

\title{Shell effects and the neutron emission within the multi-dimensional Langevin model for nuclear fission}
\author{ F.A. Ivanyuk}
\email{ivanyuk@kinr.kiev.ua}
\affiliation{Institute for Nuclear Research, 03028 Kiev, Ukraine}

\author{S.V. Radionov}
\email{radionov@kinr.kiev.ua}
\affiliation{Institute for Nuclear Research, 03028 Kiev, Ukraine}


\author{C. Ishizuka}
\email{ishizuka.c.686a@m.isct.ac.jp}
\affiliation{Institute of Science Tokyo, Tokyo, 152-8550 Japan}
\author{S. Chiba}
\email{absmithjp@gmail.com}
\affiliation{Professor Emeritus, Institute of Science Tokyo, Tokyo, 152-8550, Japan}
\date{today}

\begin{abstract}
We solve the Langevin equations for the time evolution of parameters that describe the shape of fissioning system. On each integration step, we calculate the probability of neutron emission and estimate whether a neutron would be emitted or not. If yes, we decrease the excitation energy of the nucleus by the neutron separation energy plus the average energy of the emitted neutron, switch to the layer of potential energy surface with a smaller number of neutrons and continue the process of integration. If the trajectory reaches the scission point, we check how many neutrons were emitted along this trajectory. The pre-scission neutron multiplicity $M_{pre}$ is defined by the ratio of the total number of emitted neutrons to the total number of fission trajectories. Besides $M_{pre}$, the mass distribution of fission fragments, the distribution of emitted neutrons with respect to the fission stage (deformation of system) and the distribution of emitted neutrons with respect to their energies are calculated. The calculated quantities are compared with the available experimental data.
\end{abstract}


\maketitle

\section{Introduction}
\label{intro}
The approach based on Langevin equations \cite{langevin} has been successfully applied in various branches of theoretical physics and chemistry for many years. In nuclear physics, this approach is  used for the description of fission or fusion  processes at excitations above the fission barrier \cite{nix76,wada93,froebrich,pomorski1996,karpov,asano,adeev2005,aritomo,mazurek,our17,scirep,pasha,kosenko,pomo2024,aritomo2025,song2026}.
In these works the Langevin equation was solved in 1-4 dimensions
with macroscopic \cite{werwhe,wall1,runswiat} or microscopic \cite{hofrep,hofbook,ivahof} transport coefficients.

The approach describes quite well the mass distributions and kinetic energies of fission fragments, the multiplicities of emitted neutrons, and other observable of fission or fusion  processes.

The five-dimensional calculations were published so far only in \cite{sierk,5dimen,sublead}.
Within the five-dimensional  model, several observables, such as the mean mass asymmetry seen in fission, the width of the mass yields of the heavy and light peaks, the approximate average fragment kinetic energy for fission of actinide nuclei, both spontaneous and induced by neutrons of energies of up to the threshold for second-chance fission, were accurately reproduced, \cite{sierk}.
The extension of the model to the five-dimensional shape parametrization made it also possible to reproduce such fine details of the fission process as transition from mass-symmetric to mass-asymmetric via three-humped mass distribution in thorium isotopes \cite{5dimen} and remarkably accurately reproduce the fission fragment mass distributions
 in fission of $^{180}$Hg and $^{190}$Hg at few values of the initial excitation energy, \cite{sublead}.

In the main part of the above mentioned publications, the so-called first chance fission was considered. The influence of the light particles and gamma-ray emission was neglected. However, during the evolution of the fissioning nucleus from the compound
nucleus to the scission configurations, the de-excitation of the nucleus by emitting light
particles, neutrons, protons, or $\alpha$ particles and gammas takes place. At relatively low excitation energies $E^*$ smaller than 80 MeV, only neutron evaporation takes place, while the emission of a proton or $\alpha$
particle is unlikely \cite{pomo2000}. Emission of the high-energy $\gamma$-rays in
competition with neutron evaporation is rare and is therefore neglected in the present study.

The coupling of Langevin equations with the particle emission has been investigated for a long time. We would refer here to the two recent publications \cite{pomo2024,aritomo2025}. In \cite{aritomo2025}, the authors use a three-dimensional Langevin model and take into account the evaporation by the compound nucleus before scission. The fission fragment mass distributions and the pre-scission neutron multiplicities are calculated for the isotopes of uranium, neptunium, and plutonium. Like in our present work, the authors discuss at which deformation the neutrons are emitted. In Fig.4, they present the distribution of emitted neutrons in two-dimensional collective space for fission of $^{238}$U at $E^*$=45 MeV. Unfortunately, they do not show similar results at smaller excitation energies.

Pomorski et al., \cite{pomo2024} use a four-dimensional Langevin model. Like \cite{aritomo2025}, they calculate
pre-scission emission, but they also compute the emission of neutrons by
the fragments after scission, i.e., post-scission. Like \cite{aritomo2025},
they discuss at which deformation the neutrons are emitted, but also at high $E^*$ only.

In the present work, we use the detailed description of the fission process offered by the Langevin approach to study the simultaneous shape evolution and the neutron emission. As the demonstration we present calculated results for the fission of
$^{236}$U at the excitation energies $10<E^*<40$ MeV. For this, we modify the previously used  Langevin model in the following way.

On each step of the integration of equation, we estimate whether a neutron was emitted or not. If yes, we reduce the local excitation energy by the neutron separation energy plus the average energy of the emitted neutron, switch to the deformation energy layer with smaller neutron number and continue integration of the Langevin equations. The tensors of friction and inertia are not modified since they do not depend on the excitation energy and are smooth functions of the particle number. In this way, we get very detailed information on the process of emission. We can figure out at which stage of fission the neutrons are emitted, what the number of emitted neutrons is with a given kinetic energy, and what is the number of emitted neutrons per fission event.

The computation of dN/dt takes some additional time. Besides, due to the neutron emission, the excitation energy becomes smaller; consequently, the fission probability is also smaller. These two factors increase the computation time substantially.
In order to bring it to an acceptable level, we restrict the number of collective variables to four, the neck parameter $\eps$ is kept fixed, $\eps=0.25$. That is a little bit less than the standard value $\eps=0.35$ used in our previous works. We have checked that with $\eps=0.25$ the calculated fragment mass distributions for fission of $^{236}$U are closer to the experimental results. The incorporation of neutron emission into the five-dimensional Langevin model will be the subject of future work.

To estimate whether a neutron was emitted or not, we need some approximation for the neutron emission rate dN/dt.
The popular models \cite{weiss2,dipori,pomo2002} supply the neutron emission probability for the whole fission process and do not fit well for use in the Langevin approach. The description based on the Langevin equation is very detailed. At a given moment (fission stage), the trajectory does not "know" whether it will reach the scission point, or will stay within the potential well around the ground state or will reach the limits of the deformation space and will be abandoned. The instantaneous emission rate can not depend on the fission barrier height or on the cross section of the inverse reaction. It could depend only on the local quantities like the shape of the fissioning system, or the local excitation energy, or collective velocities at a given moment.
To this end, we have developed an analytical approximation for $dN/dt$ based on the continuity equation and the Fermi-gas model.

In Section \ref{langevin}, we present the main relations of the Langevin approach used in the present work.  In Section \ref{emission} we explain how the pre-scission neutron emission is taken into account. Section \ref{results} contains the results of numerical calculations of the fission fragments mass distributions, dependence of the number of emitted neutrons on the fission stage, the pre-scission neutron spectra and the pre-scission neutron multiplicity.
 Section \ref{summa} contains a summary.
 The details of the calculations of the surface integral in the expression for the neutron emission rate dN/dt are given in Appendix A.

\section{The Langevin approach}
\label{langevin}
In the Langevin approach, one solves the set of differential equations for the time evolution of collective variables $q_{\mu}$
describing the shape of the nuclear surface. For the shape
parametrization we use in our works that of the two-center shell model (TCSM) \cite{tcsm}. In this model, the shape of the axially symmetric nucleus is characterized by 5 deformation parameters $q_{\mu} =z_0/R_0, \delta_1, \delta_2, \alpha $ and $\epsilon$.
Here, $z_0/R_0$ refers to the distance
between the centers of left and right oscillator potentials, with $R_{0}$ being the radius of the spherical nucleus. The parameters $\delta_1$ and $\delta_2$ describe the
deformation of the right and left parts of the nucleus. The fourth parameter
$\alpha $ is the mass asymmetry, and the fifth parameter $\epsilon$ of TCSM shape
parametrization regulates the neck radius. All details about the shape variables used in the present work
can be found in Ref. \cite{5dimen}. 

The first-order differential equations (Langevin equations) for the time
dependence of the collective variables $q_{\mu }$ and the conjugate momenta
$p_{\mu }$ are:

\belar{lange}
\frac{dq_\mu}{dt}&=&\left(m^{-1} \right)_{\mu \nu} p_\nu , \\
\frac{dp_\mu}{dt}&=&-\frac{\partial F(q,T)}{\partial q_\mu} - \frac{1}{2}\frac{\partial m^{-1} _{\nu \sigma} }{\partial q_\mu} p_\nu p_\sigma\nonumber\\
 &-&\gamma_{\mu \nu} m^{-1}_{\nu \sigma} p_\sigma+g_{\mu \nu} R_\nu (t),
\end{eqnarray}
where the sums over the repeated indices are assumed. In Eq.(2) $F(q, T)$ is the
temperature dependent free energy of the system, $\gamma _{\mu \nu }$
and $(m^{-1})_{\mu \nu }$ are the friction and inverse of mass tensors, and $g_{\mu \nu}R_{\nu}$(t) is the random force.

The potential energy $F(q)$ is calculated within the macroscopic-microscopic model,
\begin{equation}\label{free}
F(q)=E_{LDM}(q)+\delta F(q,T) .
\end{equation}

The macroscopic part of the energy $E_{LDM}(q)$ is calculated within the folded Yukawa model \cite{suek74,iwam76,sato79}.
At zero temperature, the shell correction to the free energy $\delta F(q, T=0)$ coincides with the shell correction to the collective potential
energy $\delta E(q)$. The shell correction $\delta E(q)$ is calculated by Strutinsky's prescription \cite{struti,brdapa} from the energies of single-particle states in the deformed Woods-Saxon potential fitted to the TCSM shapes.
The shell correction $\delta E(q)$  contains contributions from the shell effects within the independent particle (shell) model $\delta E_{shell}$  and in the pairing energy $\delta E_{pair}$ as shown in the equation
\begin{equation}
\label{deltae}
\delta E(q) =\sum_{n,p} \left( \delta E_{shell}^{(n,p)}(q) + \delta E_{pair}^{(n,p)} (q) \right) .
\end{equation}
The dependence of $\delta F(q,T)$ on the excitation energy was taken into account by the method developed in \cite{shcot}.
\belar{delffit}
\delta F(E^*)=
 \delta F(0),\quad\text{if}\quad\vert\delta F_{shell}(0)\Phi(E^*)\vert\ge \vert\delta F(0)\vert\,, \nonumber\\
 \text{or}\quad\delta F_{shell}(0)\Phi(E^*),\,\text{if}\,\vert\delta F_{shell}(0)\Phi(E^*)\vert\leq \vert\delta F(0)\vert\,.\,\,
\end{eqnarray}
The function $\Phi(E^*)$ is given by
\bel{Phi}
\Phi(E^*)=(e^{-E_1/E_0-1})/(e^{(E^*-E_1)/E_0}-1).
\end{equation}
The parameters $E_0$ and $E_1$ were defined by fitting the calculated $\delta F_{shell}(E^*)$ by the function (\ref{Phi}).
The approximation for the values of $E_0$ and $E_1$ averaged over the mass number $A$ are given in \cite{shcot},
\belar{constants}
E_0&\approx& (-467+236 A^{1/3}-38.6 A^{2/3}+2.24 A)\mev,  \nonumber\\
E_1&\approx&(-391+230 A^{1/3}-43.6 A^{2/3}+2.62 A)\mev.
\end{eqnarray}

To project the five-dimensional deformation energy onto the 2-dimensional surface, we show  the mean value
$\langle E(R_{12},\alpha)\rangle$ given by
\begin{equation}\label{eavr}
\langle E(R_{12},\alpha)\rangle=\sum E(q_i)e^{-E(q_i)/T_{coll}}\slash \sum e^{-E(q_i)/T_{coll}} .
\end{equation}
Here $q_{i}$ is the set of parameters $q_{i}\equiv \{z_0/R_0, \delta_1,\delta_2, \alpha, \eps \}$ and for $T_{coll}$ we used the value $T_{coll}$= 1 MeV. The summation in (\ref{eavr}) is carried out over the deformation points $q_i$ whose distance between right and left centers of mass does not deviate from $R_{12}$ more than $\pm 0.1 R_0.$

As one can see in Fig. \ref{edef}, the mean value of the potential energy $\langle E(R_{12},\alpha)\rangle$ is mass-symmetric before the saddle. At the saddle and above the clear mass-asymmetric valleys are seen.
\begin{figure}[ht]
\centering
\includegraphics[width=0.9\columnwidth]{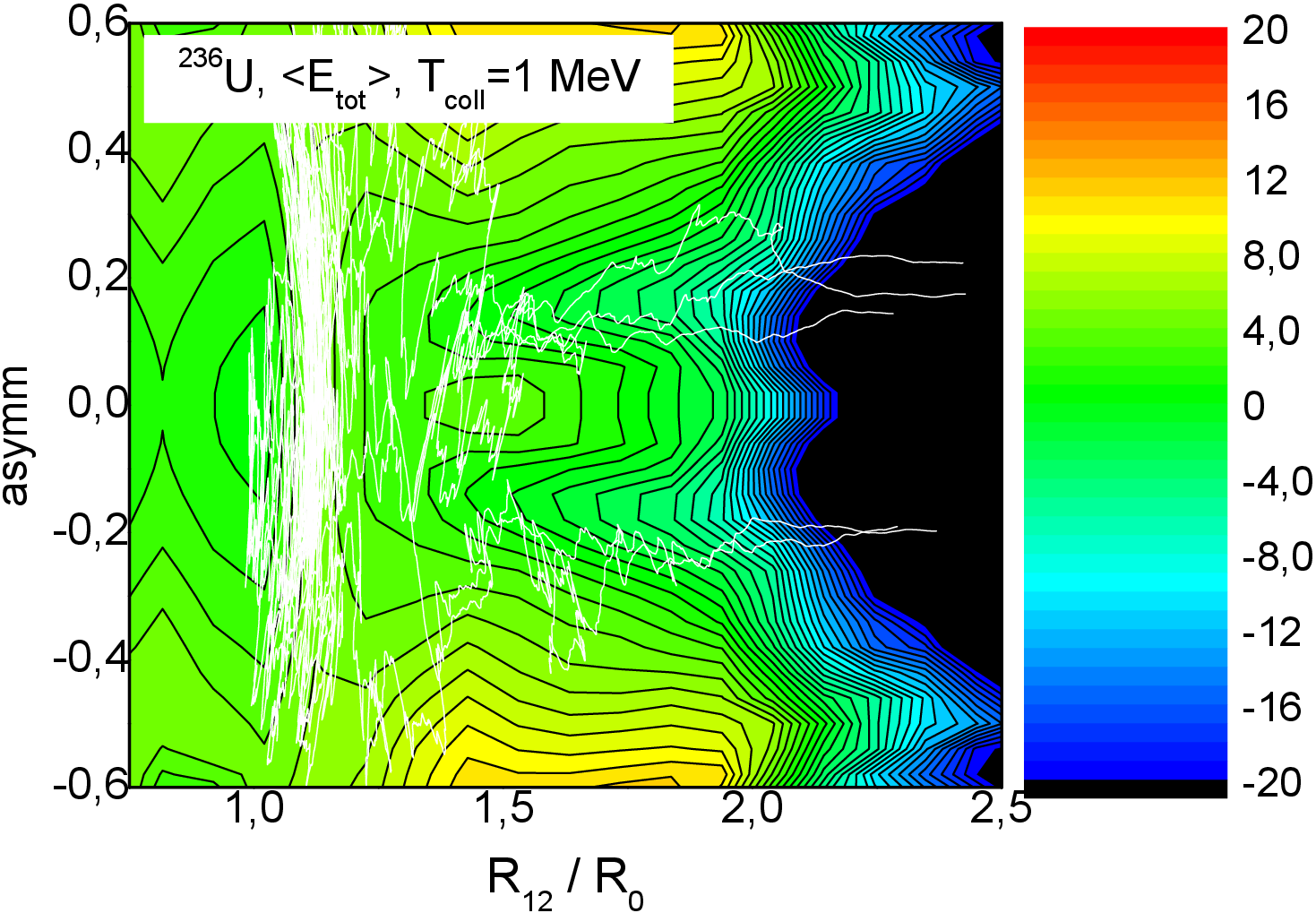}
\caption{The average value (\ref{eavr}) of the deformation energy of $^{236}$U at zero excitation energy as a function of the mass asymmetry $\alpha$ and the distance  $R_{12}$ between centers of mass of the left and right parts of nucleus. Thin white lines show few examples of fission trajectories.}
\label{edef}
\end{figure}

The compound nuclei formed in heavy-ion collisions are rotating nuclei. Their average angular momenta may be not small, and
consequently the effective fission barrier height could becomes smaller. The reduction of the fission barrier height may influence the outcome of Langevin calculations.

For the collisions with a medium-mass or heavy projectile, one can reach pretty large angular momenta even at low excitation energies. For a light projectile, or a neutron, the average value of angular momentum $L$ is not very large, and the effect of rotation on the mass distributions is very small. At the excitation energies considered in the present work, the potential energy was calculated at $L=0$.

The collective inertia tensor $m_{\mu \nu}$ is calculated within the
Werner-Wheeler approximation \cite{werwhe}
and for the friction tensor $\gamma_{\mu \nu }$ we
used the wall-and-window formula  \cite{wall1,runswiat}.

The random force $g_{\mu \nu}R_{\nu }(t)$ is the product of the normally distributed white noise
$R_{\nu}(t)$, $\langle R_{\mu}(t)R_{\nu}(t^{\pr})\rangle=2 \delta_{\mu\nu}\delta (t-t^{\pr})$, and the temperature-dependent strength factors
$g_{\mu \nu}$. The factors $g_{\mu \nu }$ are related to the temperature
and the friction tensor via the modified Einstein relation,
\begin{equation}\label{teff}
g_{\mu \sigma } g_{\sigma \nu } =T^\ast \gamma _{\mu \nu }
\,,\,\,\,\,\,\,\,\,\,\,\,\,\,\,\,\,\,\,\,\,\,\,T^\ast
=\frac{\hbar \varpi}{2}\coth \frac{\hbar \varpi}{2T}\,\,,\,\,\,\,
\end{equation}
where the effective temperature $T^{\ast }$ is defined in the references \cite{pomhof,hofkid}. The parameter $\varpi$
 represents the local frequency of collective motion \cite{hofkid}.
In the present work, we used for $\varpi$ the deformation-independent value $\hbar\varpi$=2 MeV.

Usually, the initial  values of momenta $p_{\mu}$ are set to zero, and calculations are
started from the ground state deformation.
The calculations are continued until the trajectories reach the "scission point". In present work, the "scission point" was defined as the point in deformation space where the neck radius becomes smaller than $r_{neck}^{crit}$ = 2 fm.

\section{The  neutron emission rate formula}
\label{emission}
To obtain an expression for the rate of emission of neutrons
from an atomic nucleus, we start from quantum
kinetic equation for the phase-space distribution function
$f({\bf r},{\bf p},t)$ \cite{martin59,kerman76,brink81} ,
\begin{eqnarray}
\frac{\partial }{\partial t}f({\bf r},{\bf p},t)+
\frac{1}{m}{\bf p}\cdot \nabla_{\rm r}f({\bf r},{\bf p},t)-\nonumber\\
-\frac{2}{\hbar}{\rm sin}\left(\frac{\hbar}{2}\nabla_{\rm p}^{f}
\cdot \nabla_{\rm r}^{U}\right)U({\bf r},t)f({\bf r},{\bf p},t)=0 .
\label{kineq}
\end{eqnarray}
Here, $m$ is mass of neutron, $U\equiv U({\bf r},t)$ is the
self-consistent nuclear mean field, in which all the nucleons
are moving freely; $\nabla_{\rm r}^{U}$ is the gradient operator,
acting in the coordinate space ${\bf r}$ on the function
$U({\bf r},t)$ and $\nabla_{\rm p}^{f}$ is the gradient operator,
acting in the momentum space ${\bf p}$ on the function $f({\bf r},{\bf p},t)$.

In the sequel, we will consider the case of nucleons, placed in
a potential well of the constant depth $U_0$, $U({\bf r},t)=-U_0$.
In this case, the kinetic equation (\ref{kineq}) takes the form
\begin{equation}
\frac{\partial }{\partial t}f({\bf r},{\bf p},t)+
\frac{1}{m}\left({\bf p}\cdot \nabla_{\rm r}f({\bf r},{\bf p},t)\right)=0.
\label{kineq2}
\end{equation}

Multiplying both sides of the last equation by $d{\bf p}$ and
integrating it over all possible momenta ${\bf p}$, we obtain
\begin{equation}
\frac{\partial }{\partial t}\rho_n({\bf r},t)+
\frac{1}{m}\int d{\bf p}\left({\bf p}\cdot \nabla_{\rm r}f({\bf r},{\bf p},t)\right)=0,
\label{kineq3}
\end{equation}
where
\begin{equation}\label{locdens}
\rho_n({\bf r},t)=\int d{\bf p} f({\bf r},{\bf p},t)
\end{equation}
is a local density of neutrons. The second term
in the left-hand side of Eq.~(\ref{kineq3})
may be rewritten as
\begin{equation}
\frac{1}{m}\int d{\bf p}\left({\bf p}\cdot \nabla_{\rm r}
f({\bf r},{\bf p},t)\right)=\nabla_{\rm r}{\bf j}({\bf r},t),
\label{second}
\end{equation}
where
\begin{equation}
{\bf j}({\bf r},t)=\frac{1}{m}\int d{\bf p}~{\bf p}f({\bf r},{\bf p},t)
\label{locdensi}
\end{equation}
is a local current of neutrons.

Thus, the zeroth moment (\ref{kineq3}) of the kinetic
equation (\ref{kineq2}) takes the form of a continuity
equation
\begin{equation}
\frac{d \rho_n({\bf r},t)}{dt} \equiv
\frac{\partial \rho_n({\bf r},t)}{\partial t}+
\nabla_{\rm r}{\bf j}({\bf r},t)=0.
\label{conteq}
\end{equation}
This equation simply expresses the fact that during time interval
$dt$ the amount of neutrons, escaping from a total surface of a nucleus,
$S$, is the same as the amount of neutrons, entering the surface.

Therefore, the rate of change of the number of neutrons, $N$,
escaping from the nucleus, may be represented as
\begin{equation}
\frac{d N}{dt}=-\frac{\partial }{\partial t}
\left(\int d{\bf r}\rho_n({\bf r},t)\right)=
\int d{\bf r}~\nabla_{\rm r}{\bf j}({\bf r},t),
\label{dNbdt}
\end{equation}
where the continuity equation (\ref{conteq}) was used.
The volume integral in the last equality in Eq.~(\ref{dNbdt})
may be rewritten in terms of the surface integral over $dS$,
\begin{equation}
\int d{\bf r} \left(\nabla_{\rm r}{\bf j}({\bf r},t)\right)=
\int dS \left({\bf n} \cdot {\bf j}({\bf r},t)\right),
\label{Gauss}
\end{equation}
where ${\bf n}$ is the outward pointed normal unit vector to the surface
element $dS$.

Substituting the local current of neutrons {\bf j}({\bf r},t)
(\ref{locdensi}) into Eq.~(\ref{Gauss}),
one can represent the rate of neutron emission (\ref{dNbdt}) as
\begin{eqnarray}\label{Sint0}
\frac{d N}{dt}&=&\frac{1}{m}\int dS \int d{\bf p}
\left({\bf n} \cdot {\bf p}\right)~f({\bf r},{\bf p},t) .
\end{eqnarray}

In what follows, we will use the Fermi-gas expression for the probability to find a
particle at the space point $\bf{{r}}$ with the momentum $\bf{{p}}$,
\bel{eq1}
f({\bf r},{\bf p},t)\to f(p)\equiv\frac{2}{(2\pi \hbar)^{3}}\,\frac{1}{1+\exp{[(p^{2}\,-\,p_{F}^{2})/2mT]}}.
\end{equation}
The Fermi momentum $p_{F}$ here is related to the particle number $N$ by the
condition
\bel{eq2}
\int d{\bf p}\int d{\bf r}\,f({\bf r},\,{\bf p},\,t)  \,=\,N .
\end{equation}
In case $T=$\textit{0}, the Fermi function in (\ref{eq1}) turns into $\Theta$-function,
$\Theta(\mu - p^2/2m$, $\Theta(x)$=1 for $x\ge$ 0, and is zero elsewhere.

Then the number $N$ of particles within the volume $V$ will be given by
\bel{eq3}
N=\frac{2}{(2\pi \hbar )^{3}}\,V\,4\pi \,\int\limits_0^{p_{F} } {p^{2}dp}
\,=\,\frac{2}{(2\pi \hbar )^{3}}\,\frac{4\pi Vp_{F}^{3}}{3} .
\end{equation}
From (\ref{eq3}) one gets the well-known relation
\bel{eq4}
p_{F} \,=\,\hbar (3\pi^{2}N/V)^{1/3}=\hbar (9\pi N/4R_{0}
^{3})^{1/3}\,.
\end{equation}
Below, we will need the Fermi energy - the energy of the last occupied state
\bel{EF}
E_F =p_F^2/2m=\frac{\hbar^{2}}{2mR_{0}^{2}}\left( {\frac{9\pi N}{4}} \right)^{2/3} .
\end{equation}
For the distribution (\ref{eq1}) Eq.~(\ref{Sint0}) is reduced to
\begin{eqnarray}\label{eq5}
\frac{dN}{dt}&=&\frac{1}{m}\int d  {\bf p}\, f( p)\int_{S}dS\, \Theta(\bf n \cdot \bf p)(\bf n \cdot \bf p)\nonumber\\
&=&\frac{1}{m}\int d  {\bf p}\, p f( p)\int_{S}dS\, \cos{\theta}\Theta(\cos{\theta}) ,
\end{eqnarray}
where $\theta$ is the angle between vectors $\bf p$ and $\bf n$. The $\Theta(\cos{\theta})$ in (\ref{eq5}) guarantees that the flux of particles is directed outside of the nucleus, see also \cite{friedman}.
The surface integral 
\bel{sint}
S_{int}=\int_{S}dS\,\cos\theta\Theta(\cos\theta) ,
\end{equation}
and the angular part of the integral in $\bf p$ can
be integrated analytically.

The surface integral $S_{int}$ is deformation dependent. Its value for the TCSM shape parametrization is given in Appendix A. For the spherical shape $S_{int}= \pi R_0^2$.
Note that for prolate shapes the surface area becomes larger, but the surface integral Eqs.(\ref{sint}, \ref{sint2}) - smaller.


In the present model, we consider a nucleus as a piece of infinite matter of volume $V$
put into the potential well of depth $U_0$. In this case, only particles with the kinetic energy
larger than $U_0$ can escape from nucleus, see \cite{friedman} and  Eqs. B1, B35 in \cite{pomo2000}, and $dN/dt$ turns into
\begin{equation}
\label{eq6}
\frac{dN}{dt}=S_{int}\frac{8\pi}{m(2\pi\hbar)^3}\int_{p^2/2m>U_0}
\frac{dp p^3}{1+\exp{[(p^{2}-p_F^2)/2mT]}} .
\end{equation}
Now, we introduce the notations,
\begin{equation}\label{eq7}
e\equiv -U_0+p^2/2m,~~~\mu\equiv -U_0+E_F,
\end{equation}
and with that, one obtains for $dN/dt$ in Eq.~(\ref{eq6}),
\begin{equation}\label{eq8}
\frac{dN}{dt}=S_{int}
\frac{2m}{\pi^2\hbar^3}\int_0^{\infty}de \frac{e+U_0}{1+\exp{[(e-\mu)/T]}} .
\end{equation}

The integration in $e$ in Eq.~(\ref{eq8}) is carried out over the tail of the Fermi-function,
see Fig.~\ref{fermia}.
At this tail, the Fermi-function can be approximated by the
exponent and Eq.~(\ref{eq8}) turns into
\begin{eqnarray}\label{eq9}
\frac{dN}{dt}&\approx&S_{int}\frac{2m}{\pi^2\hbar^3}\int_0^{\infty}de (e+U_0) e^{(\mu-e)/T}\nonumber\\
&=&S_{int}\frac{2m}{\pi^2\hbar^3}T(T+U_0)e^{\mu/T}\\
&=&\frac{S_{int}}{\pi R_0^2}\frac{T}{\hbar\pi}\left(\frac{9\pi N}{4}\right)^{2/3}(1+T/E_F-\mu/E_F)e^{\mu/T},\nonumber
\end{eqnarray}
where the depth $U_0$ of the potential well was expressed
in terms of the Fermi energy $E_F$ and the chemical potential
$\mu $ according to Eq.~(\ref{eq7}).
\begin{figure}[ht]
\centering
\includegraphics[width=0.8\columnwidth]{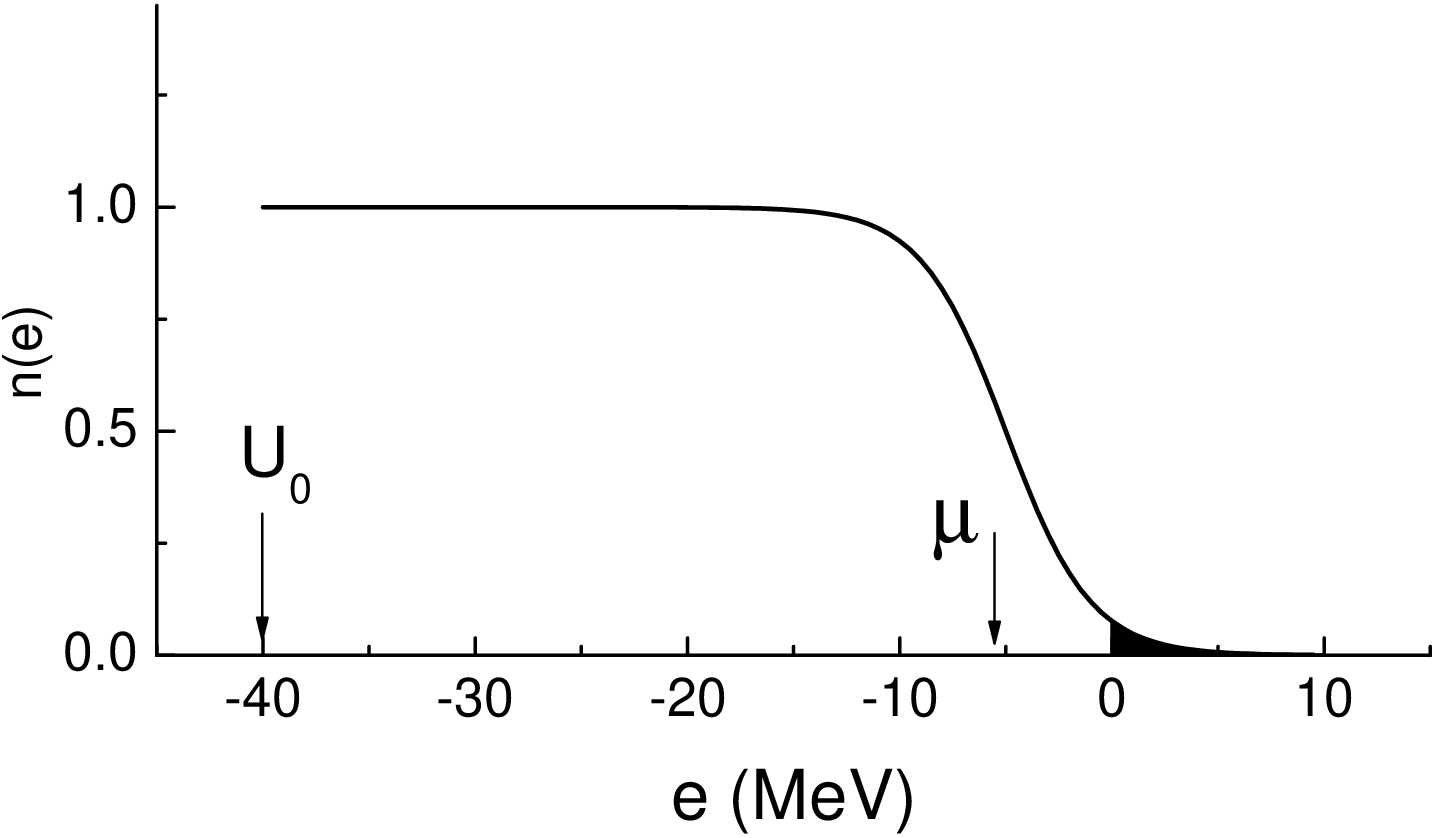}
\caption{The energy dependence of the Fermi-function
factor $n(e)=\left(1+\exp{[(e-\mu)/T]}\right)^{-1}$ in Eq.~(\ref{eq8})
for $U_0$=40 MeV, $\mu$= - 5 MeV and $T$=2 MeV.}
\label{fermia}
\end{figure}

The $E_F$ in (\ref{eq9}) is given by Eq.~(\ref{EF}), and for the chemical potential $\mu$ we take the value from the calculations of the deformation energy. The $\mu$ is the energy of the last occupied single-particle neutron state at given deformation. Thus, $\mu$ is deformation dependent.

The last line in  (\ref{eq9}) is very simple. It does not add much to the computation time and was used in the calculations presented below.

\section{Numerical results}
\label{results}
\subsection{The fission fragments mass distributions}
\label{ffmd}
We run the Langevin equations (\ref{lange}) until the scission point where the value of the neck radius becomes smaller than 2 fm.
At the scission point, we have the complete information on the deformation parameters, collective velocities and the local excitation energy. This information makes it possible to derive the distribution of fission fragments in the fragment mass or total kinetic energy.
\begin{figure}[ht]
\centering
\includegraphics[width=0.99\columnwidth]{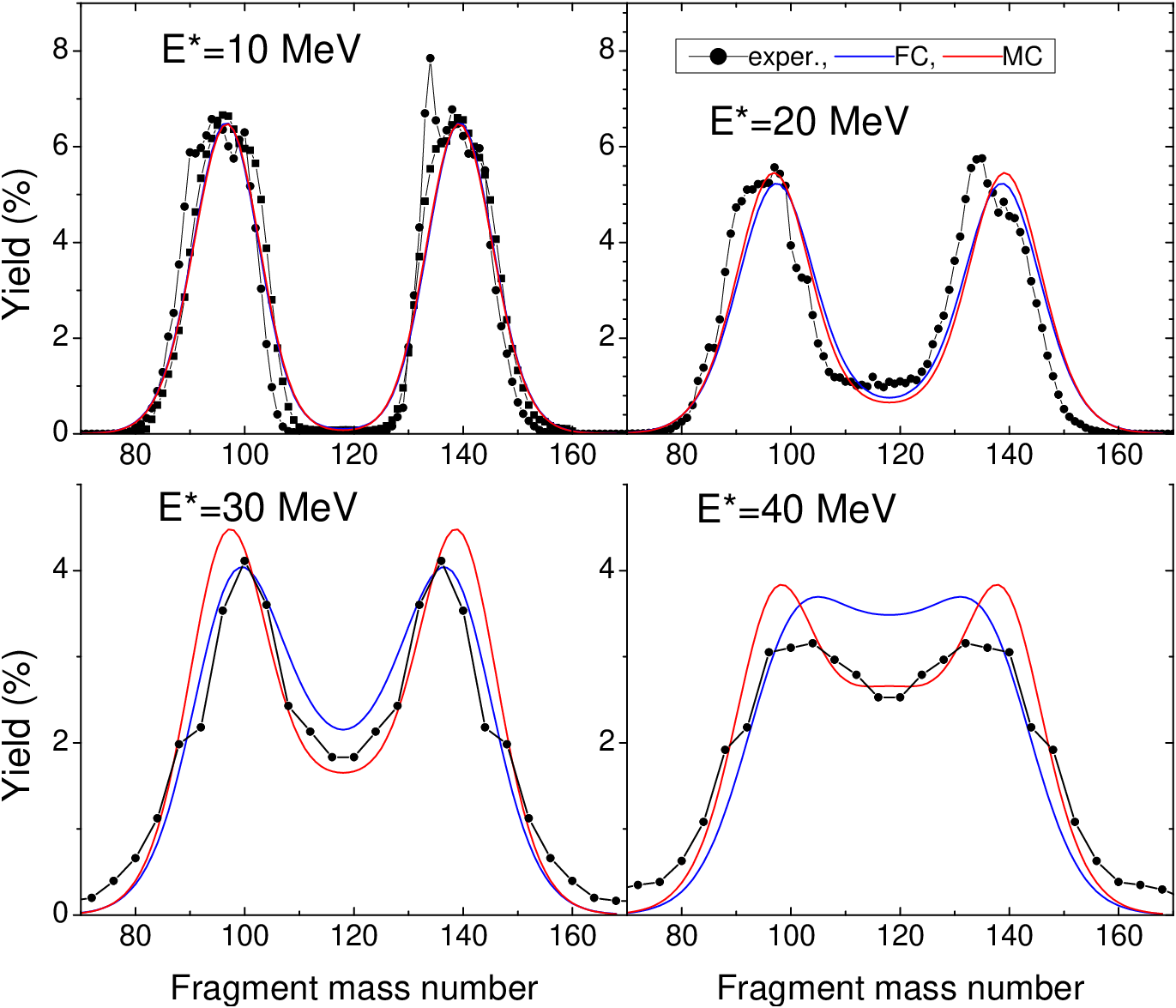}
\caption{The comparison of fission fragment mass distributions of the first chance (blue lines) and the multi-chance fission (red lines) of $^{236}$U at few values of the initial excitation energy $E^*$. The experimental mass distributions \cite{zeinalov,shibata,nishio2020} are shown by black lines with dots.}
\label{yields236}
\end{figure}

To account for the neutron emission, at each step of integration we calculate the amount of neutrons with positive energy formed during the integration step, $\Delta N = \Delta t ~dN/dt$. For $\Delta t$ we use the value $\Delta t$=0.05 fm/c, where c is the Fermi velocity.  At the same time, we generate a random number $rn$ uniformly distributed on the interval $[0,1]$. If $\Delta N$ is larger than $rn$ we believe that the neutron will be emitted.

In this case, we reduce the local excitation energy by the neutron separation energy plus its kinetic energy.
The separation energy is the energy necessary to "pull out" the neutron from the, nucleus. For the deformation energy, we use the microscopic-macroscopic approach. The least bound is the neutron on the last occupied level. This energy is called here $\mu$. The $\mu$ is negative. One has to add $-\mu$ in order to "pull out" the neutron from the nucleus, and the separation energy is $S_n=-\mu$.

It is very unlikely that the neutrons are emitted with strictly fixed energy $<e_n>$. Rather,
the distribution of neutrons in energy should be involved. For such distribution of neutrons in their velocities (or energies) we use the classical Maxwell-Boltzmann distribution,
\bel{maxv}
f(v) dv = 4\pi \left(\frac{m}{2\pi T}\right)^{3/2} v^2 \exp\left(-\frac{m v^2}{2T}\right) dv,
\end{equation}
or
\bel{maxe}
f(e) de = \frac{2}{\sqrt{\pi}}\frac{\sqrt{e}}{T^{3/2}}\exp\left(-\frac{e}{T}\right) de,\, {\rm with}\quad e\equiv\frac{mv^2}{2}.
\end{equation}

So, on each integration step, we reduce the local excitation energy by
\bel{reduce}
E^*\to E^* +\mu -T ~rm,
\end{equation}
where $rm$ is a  random number distributed according to (\ref{maxe}).

The alternative to $\mu$ could be using the neutron separation energy $S_n$ from Mollers's tables \cite{moller97}. We have checked both options. Using $\mu$ leads to somewhat better agreement of the calculated and measured pre-scission neutron multiplicities.

The comparison of fission fragment mass distributions of the first chance (blue lines) and the multi-chance fission (red lines) of $^{236}$U at a few values of the initial excitation energy $E^*$ is shown in Fig.\ref{yields236}. One can see that the account of neutron emission enhances the impact of shell effects on the yields.

For the comparison, we also show the available experimental mass distributions (black lined with dots) \cite{zeinalov,shibata,nishio2020}. One can see that the account of neutron emission substantially improves the agreement of theory and experiment. The damping of shell effects with the growing excitation energy is accurately reproduced by the approximation (\ref{delffit}).
\subsection{The distribution of emitted neutrons in the fission stage}
\label{stage}
One of the most interesting questions in nuclear fission is: at what stage of fission are the neutrons emitted? Are they emitted from the ground state region, at the fission saddle, or above the saddle? For the pre-scission neutrons, the answer to this question cannot be found by looking at the experimental data. In the experiment, the number of emitted neutrons is measured in coincidence with the fission events. It is not clear at which stage of fission the neutrons are emitted.
Here, the theory can help. Within the Langevin approach, one can see what happens to the fission trajectory at each moment. In the presented calculations, we checked at each integration step whether a neutron would be emitted.
If yes, we attach to the trajectory the information on the fission stage when it happened. Here we restrict ourselves to the elongation (center of mass distance $R_{12}$). Having the collection of data from many trajectories, we can estimate how many neutrons were emitted at a given $R_{12}$ value.

Figure \ref{distrib} shows the distribution of emitted neutrons with respect to elongation $R_{12}$ for fission of $^{236}$U at a few excitation energies, $E^*$=10, 20, 30 and 40 MeV. The orange line is the mean value of the deformation energy (\ref{eavr}) minimized with respect to $\alpha$ and calculated at $T_{coll}$=1 MeV.
\begin{figure}[ht]
\centering
\includegraphics[width=0.99\columnwidth]{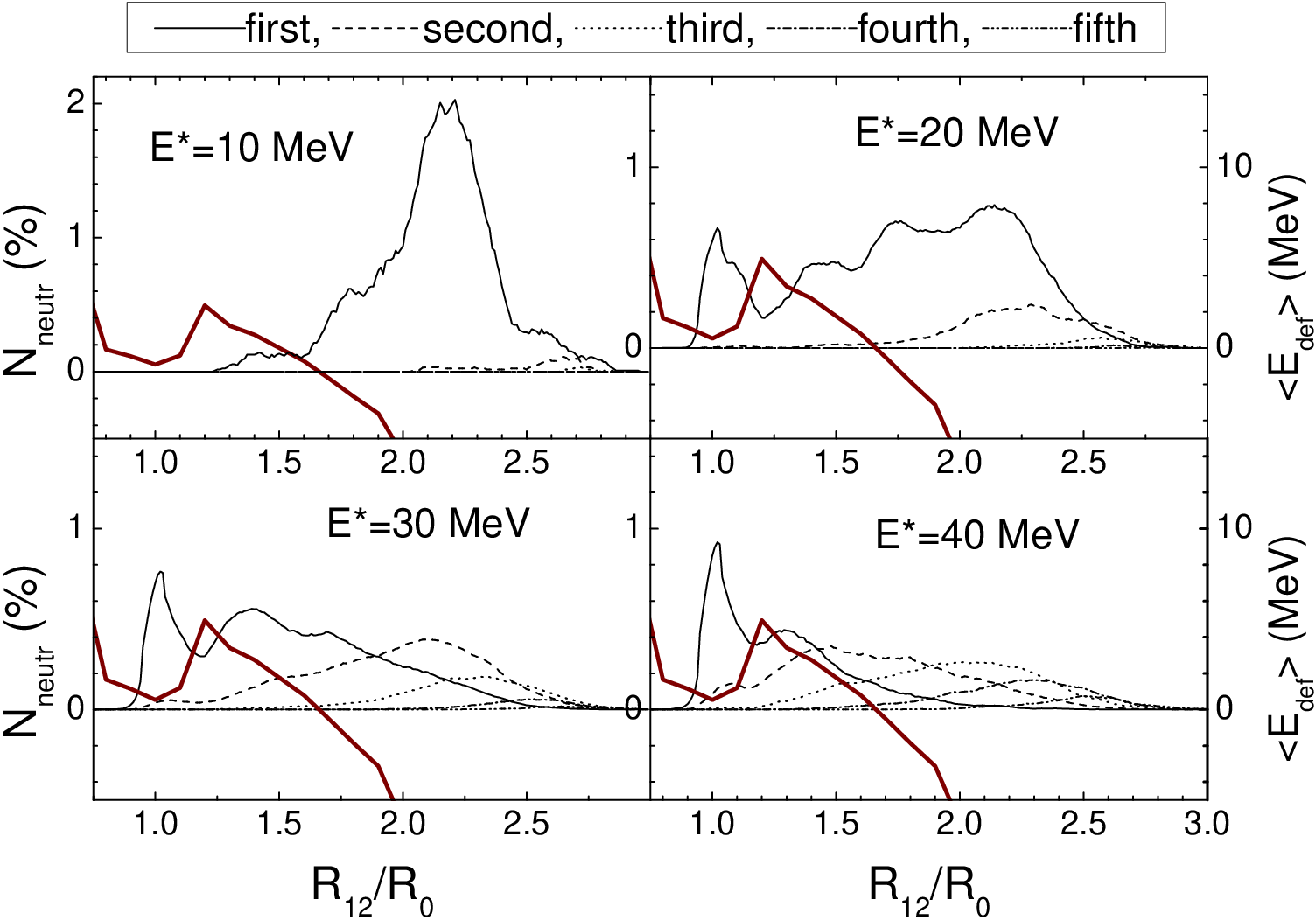}
\caption{The distribution of emitted neutrons with respect to elongation $R_{12}$ of fissioning nucleus for fission of $^{236}$U at few excitation energies, $E^*$=10, 20, 30 and 40 MeV. The orange line is the mean value (\ref{eavr}) of the deformation energy calculated with $T_{coll}$=1 MeV, right scale.}
\label{distrib}
\end{figure}

As one can see from Fig.\ref{edef}, the trajectories spent a lot of time inside the potential well before they managed to reach the saddle. One could expect that the main part of neutrons is emitted from the ground state region. But this is not the case. For $E^*$=10 MeV, nothing is emitted around the ground state. The reason is: if $E^*$=10 MeV, after neutron emission, the excitation energy gets much smaller than the fission barrier, and the trajectory can not escape the potential well. Such trajectories lead to the formation of the evaporation residue and do not contribute to fission. Only if the excitation energy is higher, 20 MeV or more, some fraction of neutrons is emitted from the ground state region. This fraction is the higher, the larger $E^*$ is. One can also notice that the number of first emitted neutrons is larger at the minimum of the potential energy and smaller at the barrier.

The second, third, ... neutrons are emitted at higher excitation energy, from outside of the fission barrier. One can expect this effect of the second, third, ... neutron emission in the mass distributions is not very large.

\subsection{The pre-scission neutron spectra}
Within the Langevin approach, the results of calculations depend crucially on the approximation (\ref{reduce}). There, we assumed a Maxwellian distribution of neutrons for each trajectory. The number of emitted neutrons summed over all trajectories is shown in Fig.\ref{spectraa} for the initial excitation energy $E^*$=10, 20, 30, and 40 MeV. Here, the comparison with the experiment would be in order. Unfortunately, the information on the pre-scission neutron emission is very scarce.

The prompt fission neutrons emitted per unit energy can be described quite accurately by a Watt distribution \cite{watt},
\bel{watt}
P(E)=Ce^{-E/a}\sinh(\sqrt{bE}),
\end{equation}
where the parameters C, a, and b are obtained by the fitting to experimental spectra, and parameter C is fixed by the normalization condition.

In Fig.~\ref{spectraa} we show by solid orange lines two examples of Watt distribution for a=T, b=1/T, C=2~$e^{-1/4}$/T$\sqrt{\pi}$  with T=0.5 MeV and T=0.75 MeV.
\begin{figure}[ht]
\centering
\includegraphics[width=0.8\columnwidth]{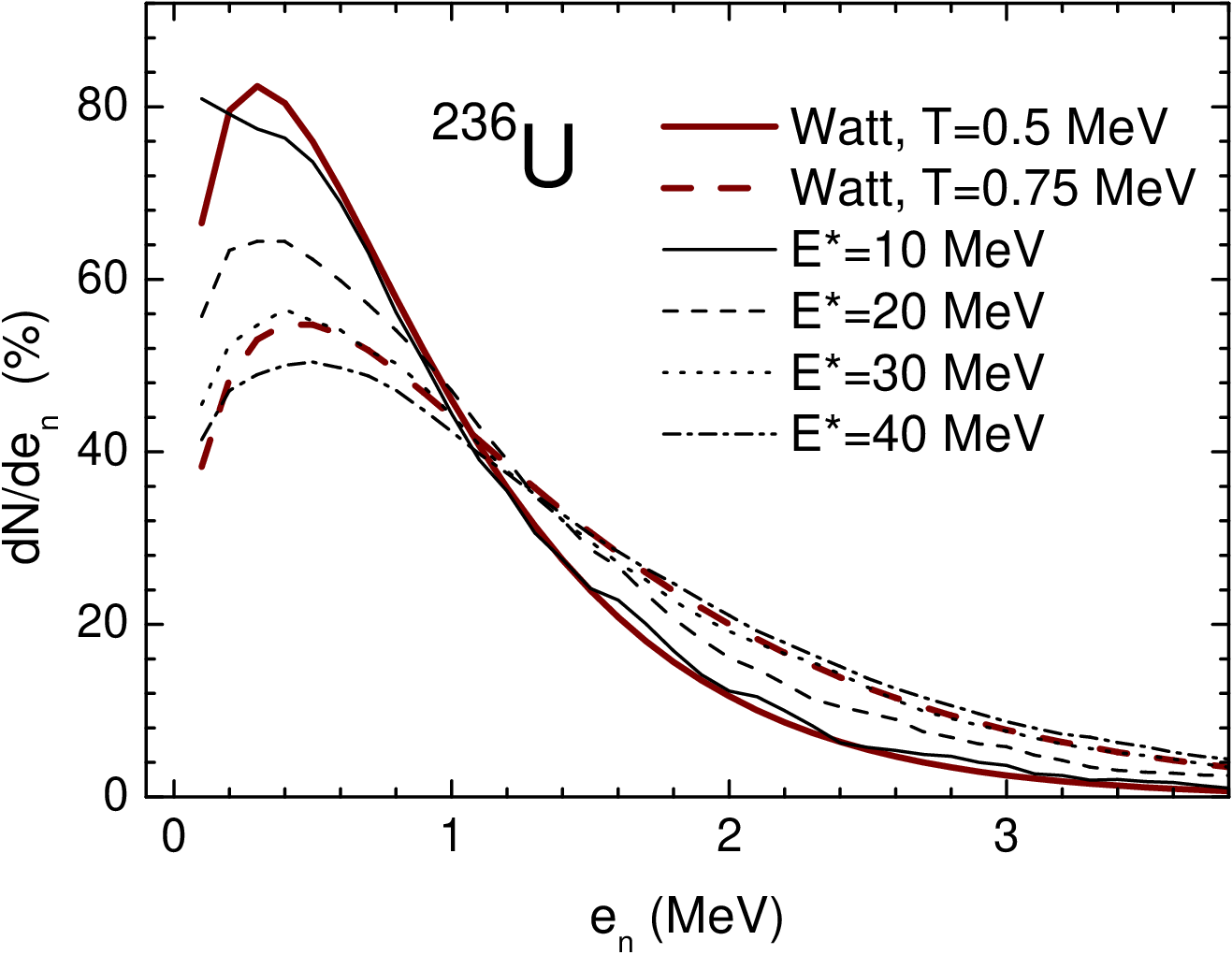}
\caption{The calculated neutron spectra for fission of $^{236}$U at few excitation energies. The Watt distributions (\ref{watt}) are shown by the orange line.}
\label{spectraa}
\end{figure}

It can be seen from Fig.~\ref{spectraa} that the shape of calculated $dN/de_n$ is very close to the Watt distributions (\ref{watt}).
\subsection{The pre-scission neutron multiplicity}
\label{multic}
The neutron multiplicity is one of the most important quantities measured in fission experiments.
The calculation of the pre-scission neutron multiplicity within the Langevin approach is rather straightforward. It does not require any additional assumptions or requirements.

At each integration step, we check whether the neutron was emitted or not. If yes, we attach to the trajectory number 1 and continue integration. If the second emission were to occur, number 1 is replaced by 2, and so on.
If the trajectory reaches the scission point, we look at how many neutrons were emitted along this trajectory, calculate the sum $N_{neutr}$ of emitted neutrons from all trajectories that have reached the scission point, and the number of trajectories $N_{fiss}$ that have reached the scission point. The ratio $N_{neutr}/N_{fiss}\equiv M_{pre}$ defines the value of the pre-scission neutron multiplicity.

In principle, so defined $M_{pre}$ depends on the number of considered trajectories $N_{fiss}$. But, if the $N_{fiss}$ becomes larger enough, $M_{pre}$ become stable against variation of $N_{fiss}$, see Fig.\ref{Mlocala}.
\begin{figure}[ht]
\centering
\includegraphics[width=0.75\columnwidth]{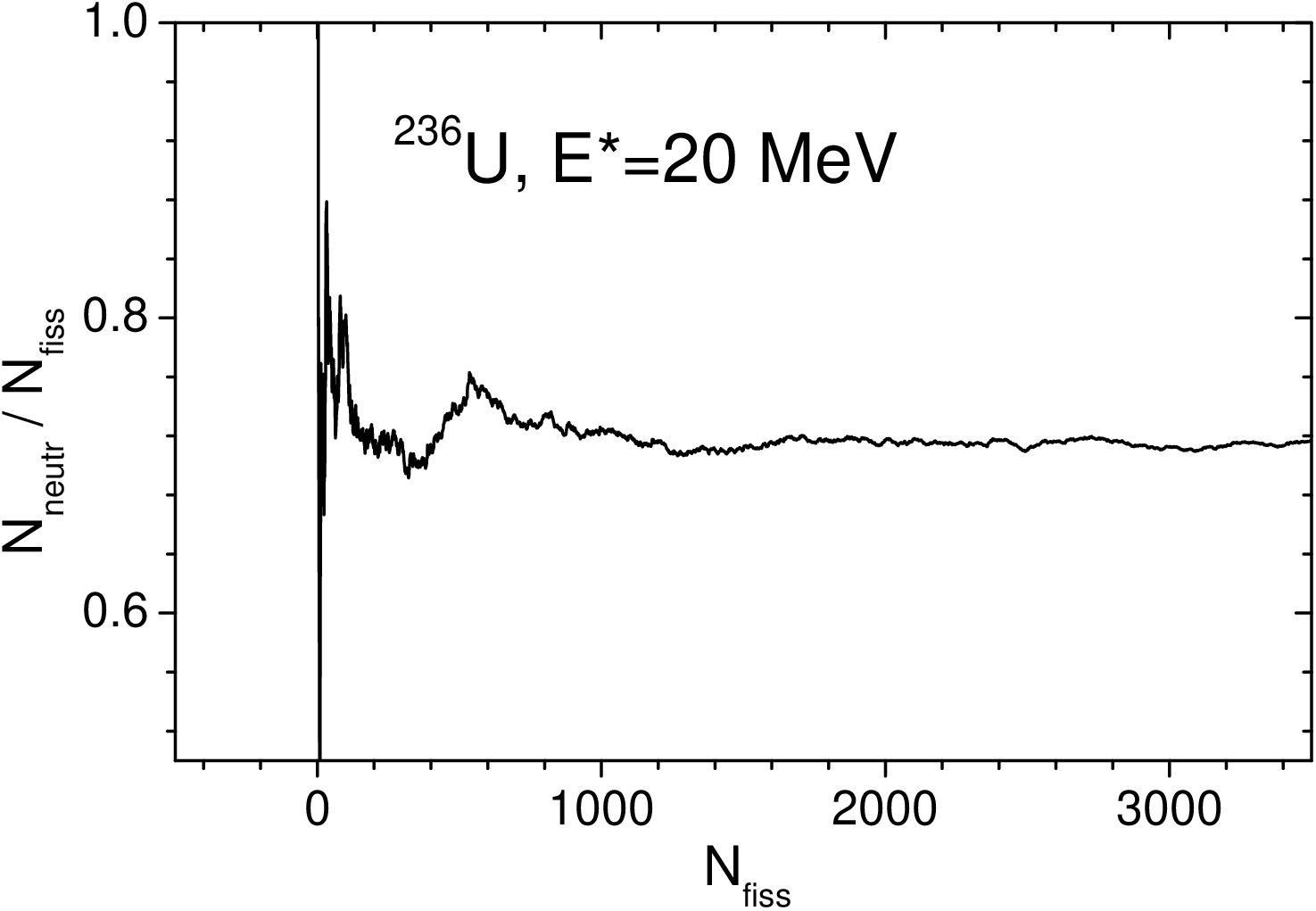}
\caption{The dependence of the pre-scission neutron multiplicity on the number of trajectories $N_{fiss}$ that reached the scission point for fission of $^{236}$U at the excitation energy $E^*$=20 MeV.}
\label{Mlocala}
\end{figure}

The pre-scission neutron multiplicity for fission of $^{236}$U is shown in Fig.\ref{Mpre} as a function of the excitation energy. Unfortunately, there is not much experimental information on the pre-scission neutron multiplicity for fission of $^{236}$U. We found only one publication \cite{kavita2024} where the pre-scission neutron emission was measured in  $\alpha+^{232}$Th reaction at $E_{lab}$= 40 and 44 MeV, green squares in Fig.\ref{Mpre}. In addition, we put the data from \cite{strecker1990} on the pre-scission neutron emission from $p+^{236}$U and $p+^{238}$U reactions.

\begin{figure}[ht]
\centering
\includegraphics[width=0.75\columnwidth]{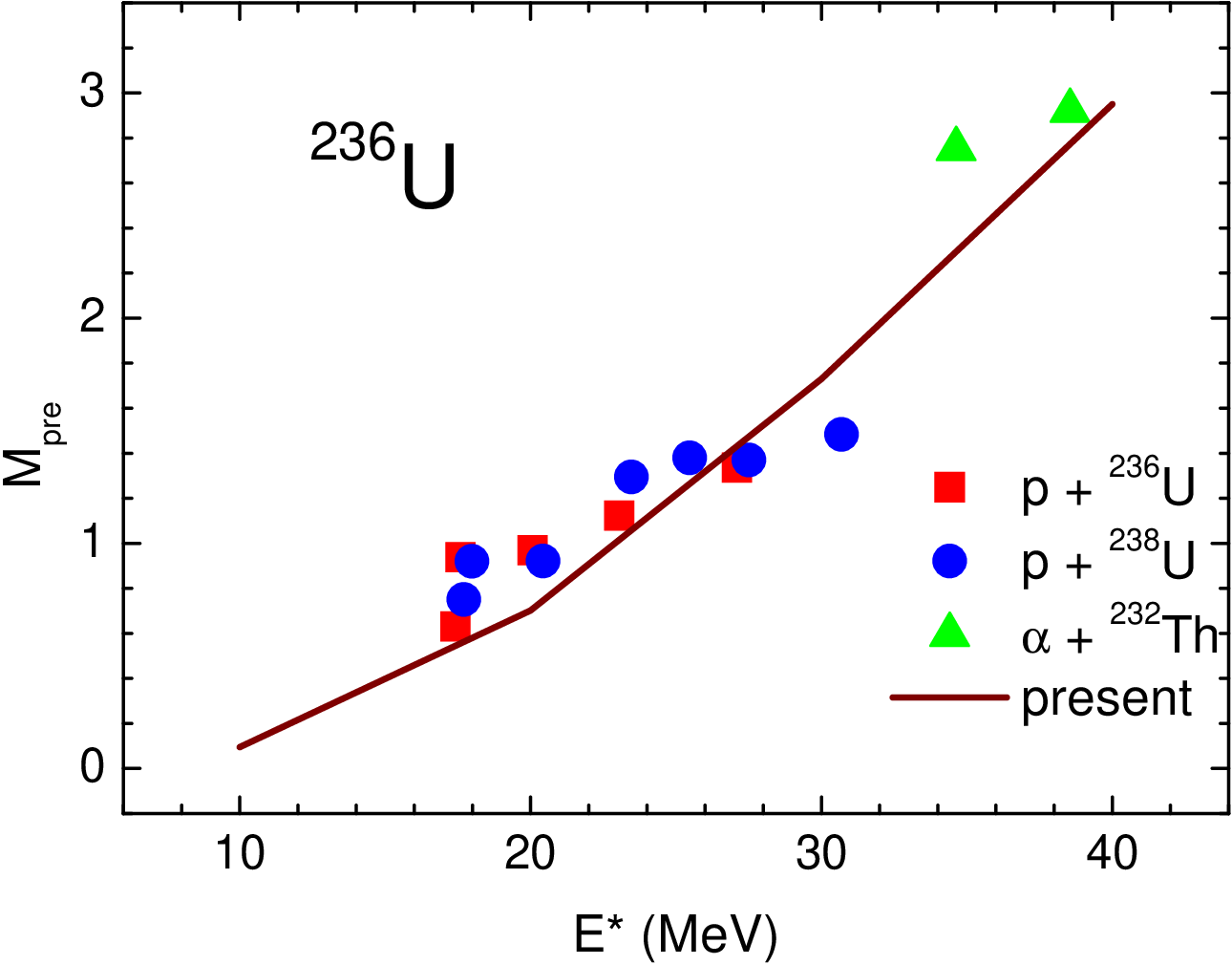}
\caption{The pre-scission neutron multiplicity  for fission of $^{236}$U at few values of the excitation energy.}
\label{Mpre}
\end{figure}
The measured and calculated $M_{pre}$ are rather close to each other. Thus, the model Eqs.(\ref{eq9}, \ref{reduce}) describes quite well both the fission fragment mass distributions and the pre-scission neutron multiplicities.
\section{Summary}
\label{summa}
We have incorporated the possibility of neutron emission into the Langevin approach developed earlier for the description of the fission process.

For the instantaneous emission rate $dN/dt$, an analytical expression was derived based on the continuity equation and the  Fermi-gas model.

The dependence of the number of emitted neutrons on the stage of fission is clarified. At $E^*$=10 MeV, all the neutrons are emitted after the system crosses the fission barrier. For a larger excitation energy, a small portion of neutrons is emitted from the potential well around the ground state. The main part of pre-scission neutrons is emitted at larger deformations, between the saddle and scission.

The calculated values of the fission fragments mass distributions and the pre-scission neutron multiplicity are in reasonable agreement with the available experimental data.

The Langevin model developed in the present work is a powerful tool for the investigation of the process of nuclear fission.
\section{Appendix A. The surface integral}
\label{sinta}

The surface integration in Eq.~(\ref{eq5}) is carried out inside the integration in momentum $\bf p$. So, for the calculation of $S_{int}$, one can choose the spatial coordinate system so that the $z$-axes, the symmetry axes of the nuclear shape, is directed along the momentum $\bf p$, see Fig.\ref{profi}.
We are interested in the flux of particles from the nucleus to the outside. That means that $p_{\perp}$, or $\cos\theta$, should be positive, as required by the $\Theta$-function in Eq.~(\ref{sint}). One can easily see from Fig. \ref{profi} that $\cos\theta$ is positive only at the part of the surface where $d\rho(z)/dz$ is negative. Note that for this part of the surface $\theta$ varies in the limits $0\leq\theta\leq \pi/2$.

To derive the expression for $\cos\theta$ we choose on profile function $\rho(z)$ a point $z_0$ and draw a touching line
\bel{touch}
\rho_t(z)=\rho_t(z_0)+\rho^{\pr}(z_0) (z-z_0)
\end{equation}
Next, we notice that the angle $\theta$ between vectors $\bf {n}$ and $\bf {p}$ is the same as the angle between the touching line and $\rho$-axes. Thus,
\bel{tantheta}
ctg\,\, \theta = \frac{\rho_t(z=0)}{z(\rho_t=0)}=\frac{\rho(z_0)-z_0\rho^{\pr}(z_0)}{z_0-\rho(z_0)/\rho^{\pr}(z_0)}=-\rho^{\pr}(z_0),
\end{equation}
and
\bel{costheta}
cos\,\, \theta = \frac{-\rho^{\pr}(z_0)}{\sqrt{1+(\rho^{\pr}(z_0))^2}}=\frac{-d\rho^2/dz_0}{\sqrt{4\rho^2(z_0)+(d\rho^2/dz_0)^2}} .
\end{equation}

Note that for $0\leq\theta\leq \pi/2$ the sign of $\cos\theta$ and $\cot\theta$ is the same. That is why the sign minus was chosen in the numerator of Eq.~(\ref{costheta}).
\begin{figure}[ht]
\centering
\includegraphics[width=0.8\columnwidth]{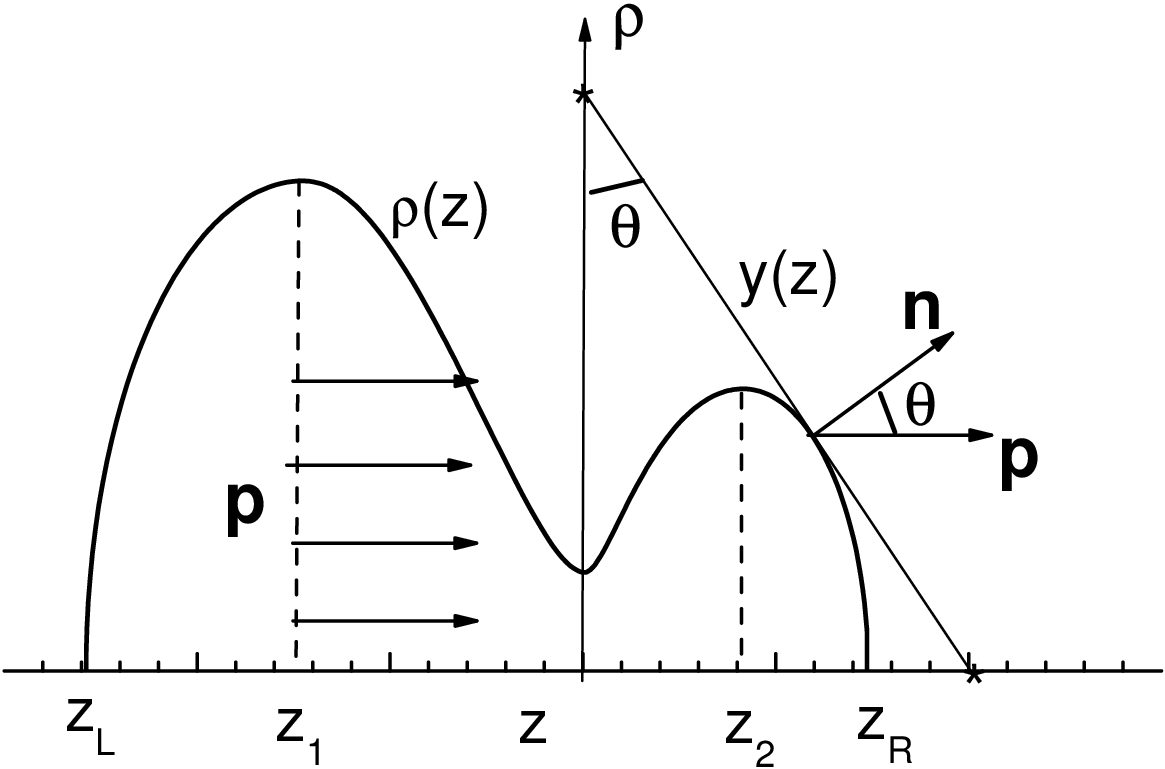}
\caption{The demonstration for the calculation of the surface integral (\ref{sint}).}
\label{profi}
\end{figure}

The surface element $dS$ is
\bel{dS}
dS=\rho(z)d\phi\sqrt{d\rho^2+dz^2}=\rho(z) \sqrt{1+(d\rho/dz)^2}d\phi dz ,
\end{equation}
and the integral (\ref{sint}) turns into
\bel{sint2}
S_{int}=-2\pi\int dz \rho(z)d\rho/dz\Theta(-d\rho/dz).
\end{equation}
Because of $\Theta$-function the integral in (\ref{sint2}) is restricted to the region where $d\rho/dz$ is negative
\belar{sint3}
S_{int}&=&-\pi\int_{z_1}^0 dz d\rho^2/dz-\pi\int_{z_2}^{z_R} dz d\rho^2/dz\nonumber\\
&=&\pi[\rho^2(z_1)+\rho^2(z_2)-\rho^2(0)].
\end{eqnarray}
For the shapes without neck, $z_2-z_1\to 0$, $\rho^2(z_1)=\rho^2(z_2)=\rho^2(0)$ and
\bel{sint4}
S_{int}=\pi\rho^2(0).
\end{equation}
For the spherical shape $S_{int}= \pi R_0^2$.
 Note that for prolate shapes the surface area becomes larger, but the surface integral Eqs.(\ref{sint}, \ref{sint2}) - smaller.

\begin{acknowledgments}
{\bf Acknowledgments.} The authors would like to express their gratitude to Prof. K. Pomorski and Dr. C. Schmitt for valuable comments and suggestions, and Dr. K. Nishio for sharing the experimental data on the fission fragment mass distributions for $^{236}$U.
\end{acknowledgments}

\end{document}